\documentclass[12pt,a4paper]{elsarticle}
\biboptions{longnamesfirst}

\journal{Theoretical Population Biology}

\usepackage[utf8]{inputenc}
\usepackage{amsmath}
\usepackage{amsfonts}
\usepackage{amssymb}
\usepackage{graphicx}
\usepackage{color}
\usepackage{url}
\usepackage{algpseudocode}
\usepackage{mathtools}
\usepackage[skip=10pt plus1pt, indent=10pt]{parskip}

\newcommand{\LOC}[0]{\ell}

\newcommand{\PHY}[0]{\mathbf{\tau}}
\newcommand{\HAB}[0]{\mathcal{A}}
\newcommand{\SFV}{$\Lambda$V}

\newcommand{\BDBD}{BD$^2$}
\newcommand{\Ex}{\mathrm{E}}

\begin{document}

\begin{frontmatter}

\title{On the connections between the spatial Lambda-Fleming-Viot model and other processes for analysing geo-referenced genetic data}
\author[1]{Johannes Wirtz}
\ead{jwirtz@lirmm.fr}
\author[1]{St\'ephane Guindon\corref{corresponding}}
\ead{guindon@lirmm.fr}
\cortext[corresponding]{Corresponding author}
\affiliation[1]{Laboratoire d'Informatique, de Robotique et de Microélectronique de Montpellier. CNRS - UMR 5506. Montpellier, France}

\begin{abstract}
    The introduction of the spatial Lambda-Fleming-Viot model (\SFV) in population genetics was mainly driven by the pioneering work of Alison Etheridge, in collaboration with Nick Barton and Amandine Véber about ten years ago \cite{barton2010,barton2013}. The \SFV~model provides a sound mathematical framework for describing the evolution of a population of related individuals along a spatial continuum. It alleviates the ‘‘pain in the torus" issue with Wright and Malécot's isolation by distance model and is sampling consistent, making it a tool of choice for statistical inference. Yet, little is known about the potential connections between the \SFV~and other stochastic processes generating trees and the spatial coordinates along the corresponding lineages. This work focuses on a version of the \SFV~whereby lineages move infinitely rapidly over infinitely small distances. Using simulations, we show that the induced \SFV~tree-generating process is well approximated by a birth-death model. Our results also indicate that Brownian motions modelling the movements of lineages along birth-death trees do not generally provide a good approximation of the \SFV~due to habitat boundaries effects that play an increasingly important role in the long run. Finally, we describe efficient algorithms for fast simulation of the backward and forward in time versions of the \SFV~model.
    \section*{Highlights}
\begin{itemize}
   \item Birth and death of lineages in the spatial Lambda-Fleming-Viot model converge to independent Poisson processes.
   \item Tree-generating processes induced by the spatial Lambda-Fleming-Viot and (super-)critical birth-death processes are equivalent in the limit of low spatial variance.
   \item This equivalence does not carry over when accounting for spatial information.  
\end{itemize}
    \end{abstract}

\begin{keyword}
Spatial Lambda-Fleming-Viot\sep Birth-Death processes\sep Duality\sep Efficient simulation
\end{keyword}

\end{frontmatter}


\section{Introduction}

The integrated analysis of genetic and spatial data in the fields of phylogeography or spatial population genetics is central to our understanding of the forces driving the evolution of living organisms in space and time. Indeed, accommodating for the evolutionary relationships between individuals of the same population or between distantly related species when analysing their spatial distribution permits the reconstruction of ancestral migration and dispersal events. It then becomes possible to examine the links between these events and past environmental or ecological changes so as to decipher the mechanisms underlying key biological processes such as speciation or the impact of natural selection on spatial patterns of biodiversity. Combining the horizontal (spatial) and vertical (evolutionary) dimensions of geo-referenced genetic data is therefore paramount in order to elucidate the mechanisms  and test  hypotheses about the underlying data generating processes.

In population genetics, Wright's island model \cite{wright1931} was the first of a series of ``migration-matrix'' models that aimed at describing the evolution of a population that is spatially structured in distinct demes (see \citet{rousset2001} for a review). Despite their relative simplicity, the island model and its descendants, including most notably the stepping stone model \cite{kimura1953}, provided population geneticists with a rich set of tools to test important biological hypotheses such as panmixia or the existence of past and/or ongoing migrations between sub-populations. 

The assumption of discrete demes is convenient mathematically. The ability to accommodate for populations that are spatially distributed along a continuum is a natural extension of the discrete assumption. That extension is expected to significantly expand the range of applications and, in numerous instances, enhance the relevance of spatial population genetics models \cite{bradburd2019}. Over the last eight decades, progresses in the development of these models turned out to be rather slow and faced serious difficulties in some cases. The isolation by distance model proposed by Sewall Wright \cite{wright1943} and Gustave Mal\'ecot \cite{malecot1948}, for instance, was shown to suffer from pathological behaviour in the long run (the so-called ‘‘pain in the torus" described by Joseph Felsenstein, \cite{felsenstein1975}), forcing population geneticists to rely on the discrete approximation aforementioned. The approaches proposed by \citet{wilkins2002,wilkins2004} addressed the ``clumping" issue that hampered the isolation by distance model. Yet, as pointed by Alison Etheridge and colleagues, the models proposed here lacked sampling consistency, implying that the time to coalescence of lineages depended on the size of the sample considered \cite{barton2010}, thereby limiting their application in practise.

While there are relevant approaches available that provide graphical summaries of populations distributed along a spatial continuum  (see e.g., \cite{novembre2008,wang2012,bradburd2016,bradburd2019}), sound mechanistic models that accommodate for continuous diffusion of individuals in their habitat along with genetic drift are scarce. In a pioneering work, Alison Etheridge, Nick Barton and Amandine Véber \cite{barton2010,barton2013} introduced the spatial Lambda-Fleming-Viot model (noted as \SFV~in the following) in an attempt to fill this gap. To the best of our knowledge, the \SFV~is the sole mechanistic model that (1) accommodates for populations distributed along a spatial continuum, under a stationary regime (i.e., the population density does not change, on average, during the course of evolution) and (2) provides a coherent account of the forward in time  evolution of a population along with a dual description of the backward in time evolutionary dynamics of a sample from that population and (3) is amenable to parameter inference using a Bayesian approach, applicable to small to moderate size data sets, e.g., see \cite{guindon2016,joseph2016}).

The properties of the \SFV~model and some extensions are well characterised mathematically \cite{veber2015,biswas2021,louvet2023}. Yet, relatively little is known about the relationships between \SFV~and other popular population genetics models. Shedding light on potential connections between these models would help delineate conditions in which the \SFV~may be well approximated by other processes, potentially leading to more efficient parameter estimation procedures. More importantly, establishing such bridges would help gain a better understanding of the biological relevance of the \SFV~process. 

In this study, we consider the non-trivial case where the rate of reproduction and extinction (REX) events in the \SFV~model is large and the radius of each event (i.e., the parent-to-offspring distance) is small. We first focus on the tree-generating process that derives from the \SFV~model forward in time in these particular conditions and show that the distribution of trees deriving from the \SFV~is well approximated by that obtained from a birth and death (BD) process. We then incorporate the spatial component in our analyses and show how the \SFV~model compares to the birth and death model with spatial coordinates fluctuating along lineages according to a Brownian process, as introduced in \cite{lemey2010} and available in the popular software package BEAST \cite{drummond2007,suchard2018}. Results from simulations indicate that habitat border effects that come into play with the \SFV~model but are ignored by the Brownian process, preclude the convergence of both models to the same process. Finally, we describe two algorithms for efficient simulation of the \SFV~process forward and backward in time, which are at the core of some of the model comparisons performed here.

\subsection{Notation and models}

We first introduce some notation that will be used throughout the manuscript. Let $n$ be the number of sampled lineages. $\PHY$ denotes a ranked tree topology with $n$ tips and $t$, the corresponding vector of $2n-1$ node times, which are defined relative to the sampling time. Throughout this study, sampling of lineages takes place at a single point in time (i.e., we do not account for heterochronous data) taken to be equal to 0. $\LOC$ is the vector of $2n-1$ spatial coordinates at all nodes in the tree.

\subsubsection{Individual-based \SFV~model}

We consider the forward-in-time version of this process here, taking place on a $w \times h$ rectangle, denoted as $\HAB$ in what follows. Individuals that constitute the population of interest are distributed uniformly at random with density $\rho$ on that rectangle. Lineage reproduction and extinction (REX) events occur at rate $\xi$, the per unit space rate. When one such event takes place, (1) individuals die with probability $\upsilon \exp(-d^2/2\theta^2)$, where $d$ is the Euclidean distance between the corresponding individual position and the location of the center of the REX event, (2) offspring are generated according to a non-homogeneous Poisson process with intensity $\rho \upsilon \exp(-d^2/2\theta^2)$ and (3) one parent for the newly generated offspring is chosen where a parent at distance $d$ from the centre has probability proportional to $\exp(-d^2/2\theta^2)$ to be selected (individuals that die on that event may also be selected as parent).

A closed-form formula for the likelihood, i.e., the joint probability density of $\tau$ and $t$ conditioned on $\upsilon$, $\theta$, $\xi$ and $\rho$ is not available. Yet, obtaining random draws from the corresponding distribution is relatively straightforward. In particular, in a manner similar to the Wright-Fisher model and Kingman's coalescent \cite{kingman1982}, the \SFV~has a backward in time dual of the forward in time process that allows for rapid simulations of  genealogies of a sample of $n$ lineages (see \cite{barton2010} and  section \ref{sec:fast_back_sim} for a description of an efficient backward in time algorithm for simulating a two-tip genealogy under the \SFV).

\subsubsection{Birth and death process with Brownian diffusion}

Beside the \SFV~model, this study focuses on the homogeneous BD model with complete sampling. According to this process, a first lineage arises at time $t_{\text{or}}$, the time of origin of the process. The rate at which any given lineage splits/dies is governed by the birth and death parameters $\lambda$ and $\mu$ respectively, i.e., the process is homogeneous so that per-lineage birth and death rates are fixed throughout. Data collection takes place in the future compared to $t_{\text{or}}$, at which point we condition the genealogy $\tau$ on having $n$ live lineages, which are all included in our sample. 

Spatial coordinates evolve in a two dimensional space, with movements along the northings considered as independent from that along the eastings. In each dimension, the spatial position of a lineage fluctuates according to a Brownian process with diffusion parameter $\sigma$. Hence, the distribution of the position at the end of a branch of length $t$ (in calendar time units) is Gaussian with mean given by the lineage position at the start of that branch and variance $\sigma^2 t$.

The very idea of using  Brownian diffusion to model the evolution of locations along a genealogy was introduced in \cite{lemmon2008,lemey2010}. Although \citet{lemey2010} focused on a ``relaxed'' version of this approach, whereby each branch in the phylogeny has its own spatial diffusion parameter, we focus here instead on the ``strict'' version of the model, with a single diffusion parameter applying to all edges of the tree. This model is noted \BDBD~in the following for BD model combined with Brownian Diffusion.
 
\section{Connecting the two models}

Our study first focuses on the comparison between between $p_{\text{BD}}(\PHY,t|\lambda,\mu,n,\sigma)$, the likelihood of the BD model, and the equivalent density for the \SFV~model, $p_{\text{\SFV}}(\PHY,t|\xi,\theta,n,\upsilon,\rho)$, when focusing only on the tree-generating parts of both models. We then examine
the link between $p_{\text{\BDBD}}(\PHY,t,\LOC|\lambda,\mu$,$n,\sigma)$  and $p_{\text{\SFV}}(\PHY,t,\LOC|\xi$, 
$\theta$,$n,\upsilon,\rho)$, the full likelihoods, i.e., including both the tree and the spatial components, of \SFV~and \BDBD. We consider the particular case where $\xi \to \infty$ and $\theta \to 0$, i.e., REX events occur at a high rate and each of them has a very small radius. We assume here that $\xi\theta^2 \to c$ for some $c\in\mathbb{R}$. Disregarding the spatial component, we refer to the ``limit'' model as \SFV$^*$.


\subsection{Birth and death rates under \SFV}
 We first focus on the rate $\mu_{\text{\SFV}}(x)$ at which an individual at position $x\in\HAB$ dies in a REX event under \SFV. Events occur uniformly on $\HAB$ at rate $\xi$, and given an event location $z\in\HAB$, the probability that an individual at position $x$ dies due to the event is $\upsilon \exp(-d^2/2\theta^2)$, where $d=\|x-z\|$. So the overall rate at which an individual at $x$ dies is obtained by integrating this probability over the habitat, multiplied by $\xi |\mathcal{A}|$, where $|\mathcal{A}|$ is the area of the habitat. So we have:
\begin{equation}
 \mu_{\text{\SFV}}(x)=\xi |\mathcal{A}|\int_{\HAB}\frac{\upsilon}{|\mathcal{A}|} \exp(-\|x-z\|^2/2\theta^2)\mathrm{d}z
\end{equation}
When $\theta \to 0$, $\xi \to \infty$ and $\theta^2 \xi \to c$, the right-hand side can be written as
\begin{eqnarray}
 \mu_{\text{\SFV}^*}(x) = 2 \pi c \upsilon,
\end{eqnarray}
We observe that the rate hence obtained does not depend on the individual's position $x$. In particular, this rate does not depend on the distance to the edges of the habitat. Therefore, in the \SFV$^*$ all individuals have a unique ``death rate" $\mu_{\text{\SFV}^*} = 2 \pi c \upsilon$. 

Similarly, individuals have a rate of ``giving birth" analogous to the birth rate of a BD process. At any REX event, the cumulative intensity of births on $\HAB$ is
\begin{equation}
\int_{\HAB} \rho \upsilon \exp(-\|y-z\|^2/2\theta^2)\mathrm{d}z \to 2 \pi \theta^2\rho \upsilon 
\end{equation}
where $z$ denotes the event location. Therefore, in the \SFV$^*$ process, the number of individuals born in one event is Poisson with parameter $2 \pi \theta^2\rho \upsilon$ and the probability that in one event $k$ individuals are born is 
\begin{equation}\label{eq:pk}
p_k:=\frac{\left(2 \pi \theta^2\rho \upsilon\right)^k}{k!}\exp(-2 \pi \theta^2\rho \upsilon)
\end{equation}
In \cite{barton2013}, it is shown that the probability that an individual at location $x$ is chosen as the parent by an event located at $z$ is 
\begin{equation}\label{eq:barton}
\frac{1}{2\pi\theta^2\rho\upsilon}\exp(-\|x-z\|^2/2\theta^2)\cdot(1+\mathcal{O}(\rho^{-1}))
\end{equation} 
We shall assume that $\rho$ is large enough such that the order term in \eqref{eq:barton} becomes negligible. However, we note that when simulating we observed that even for values of $\rho$ around $1$ this seemed to be the case on average. Combining \eqref{eq:barton} and \eqref{eq:pk}, we can calculate the rate at which an individual at $x$ is chosen as the parent by a REX event and $k$ individuals are being generated by that event as
\begin{equation}
\lambda^{(k)}_{\text{\SFV}}(x)=\xi|\mathcal{A}|\int_{\HAB}p_k\cdot\frac{1}{2\pi\theta^2\rho\upsilon|\mathcal{A}|}\exp(-\|x-z\|^2/2\theta^2)  \mathrm{d}z 
\end{equation}
Now, letting $\xi\rightarrow \infty$ tend to infinity and $\theta\to 0$ in the same way as before,  we have

\begin{equation}
\lim \lambda^{(1)}_{\text{\SFV}}(x)=\lim \frac{\xi }{\rho\upsilon}  p_1 = 2\pi c
\end{equation}

\noindent and $\lambda^{(k)}_{\text{\SFV}}(x) \rightarrow 0$ for all $k>1$. The limit thus eliminates the possibility of multiple offspring during one event, ensuring that an individual can give birth to at most one child at a time. The rate at which lineages split in the \SFV$^*$ is then
\begin{equation}
\lambda_{\text{\SFV}^*}(x)=2 \pi c 
\end{equation}


In the standard BD model, birth and death events never happen at the very same point in time, let alone along the same lineage. In the \SFV~model however, a REX event may involve the splitting of the parental lineage {\em and} the death of that same lineage. Given a REX event with centre $z$, the probability for a given lineage located at $x$ to die or to give birth to new lineages is proportional to $\exp{\left( -\|x-z\|^2/2\theta^2\right)}$. The probability of both events (birth and death) taking place is thus proportional to $\exp{\left( -\|x-z\|^2/\theta^2\right)}$. Birth or death events alone therefore become infinitely more probable than simultaneous birth and death events when the radius tends to zero so that, in that respect, the \SFV~behaves in a manner similar to the BD model.





We conclude that in the \SFV, for diminishing values of $\theta$ and increasing $\xi$, the number of offspring lineages is stochastically similar to the number of lineages in a birth and death process with $\lambda=2\pi c$, $\mu=2 \pi c \upsilon$, where $c = \theta ^ 2 \xi$ is constant. Since $0<\upsilon\leq 1$, we always have $\lambda\geq\mu$, so the process is supercritical, except in the case where $\upsilon=1$. We shall make use of the latter assumption throughout this manuscript. 


We confirmed these observations by simulating the \SFV~forward in time. At the beginning of each simulation run, we randomly selected one individual within the population. Simulations stopped whenever this individual was the target of an event, and the time at which that event took place was recorded. Three types of events can be observed: 1) The death of the individual; 2) the individual giving birth to one or more offspring individuals; and 3) death and birth of that individual at the same time. The values for $\theta$ and $\xi$ were chosen in such a way that $c=\theta^2\xi=1$ throughout our simulations. Also we set $\upsilon=1$, $\rho=20$, $w=10$ and $h=10$.

As $\theta$ decreases (and $\xi$ increases), we observe that events of the third type (simultaneous birth and death of the same lineage) become increasingly rare, while events of type one (death) and two (birth) occur at about the same frequency. For example, for $\theta=\xi=1$ the frequency of type three events observed is close to $0.08$, rises to $0.15$ for $\theta=0.25,\xi=16$, then drops to $0.01$ for $\theta=0.05$ and to effectively zero for smaller $\theta$. For type one, the frequencies are $0.74$, $0.63$, $0.54$ and $0.50$, whereas the frequencies of events of type three are $0.18$, $0.22$, $0.45$ and finally about $0.50$ as well. 

\begin{figure}
    \includegraphics[width=\linewidth]{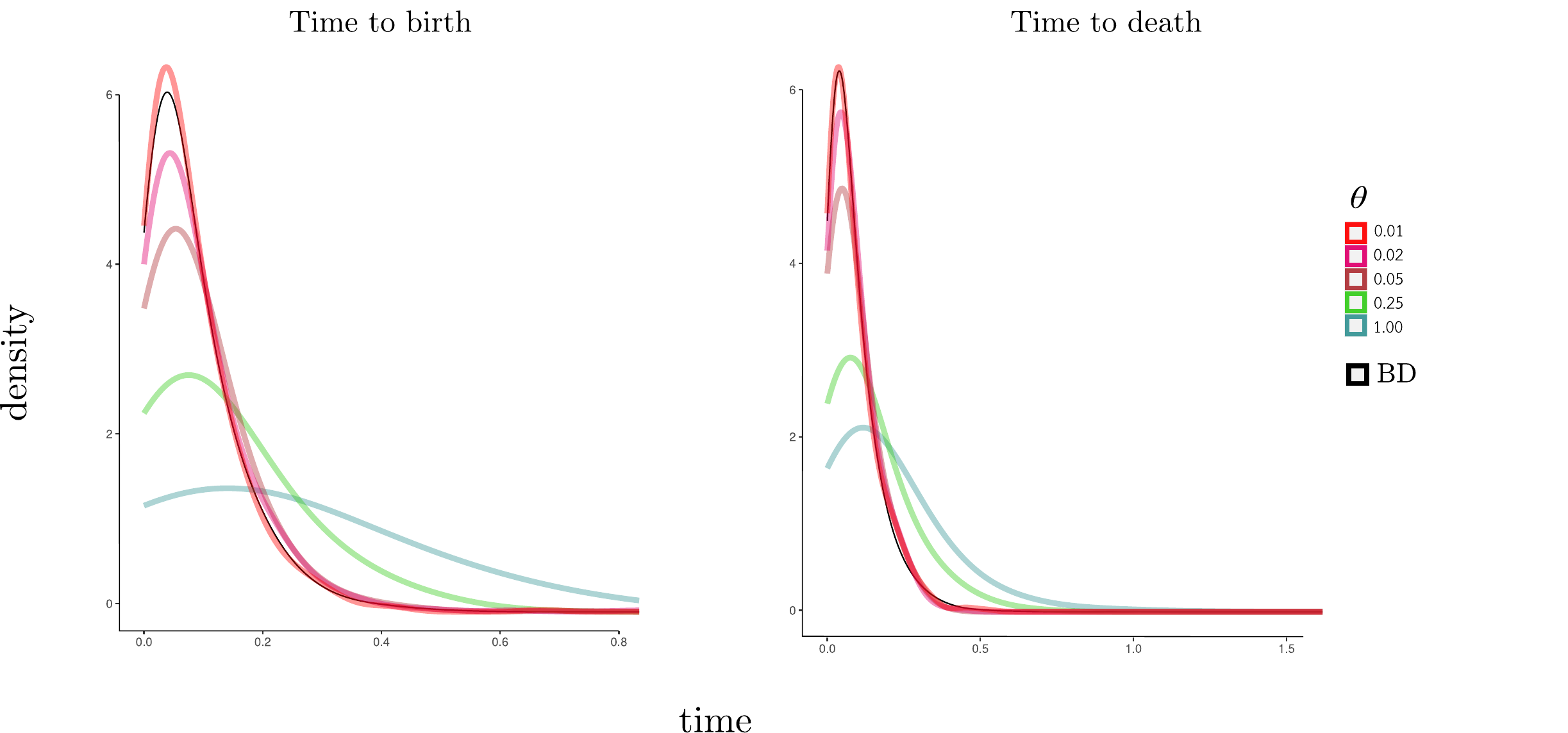}
    \caption{{\bf Densities of the times until a given individual is subject to a birth event (left) and to a death event in the \SFV.} Tested values of $\theta$ are given to the right. $\xi$ is such that $\theta^2\xi=1$. The respective densities in the BD are shown in black.}
    \label{fig:birth_death_1l}
\end{figure}

From the times recorded at which these events take place, we reconstruct the probability density of the time to an event of the respective type. These densities are represented in  Figure \ref{fig:birth_death_1l} for various values of $\theta$, while the black curves represents the densities derived for death events (right) and birth events (left), respectively in a BD process with parameter $\lambda=\mu=2\pi$, which both conform to an exponential density with parameter $\lambda=2\pi$. For decreasing $\theta$, we observe a trend of the densities in the \SFV~to approach those in the BD.
   
\subsection{On the offspring number distribution in the two processes}

In this section, we examine the distribution of the number of descendant lineages resulting from each individual in the initial population with respect to time. More specifically, we monitor the number of live descendants of every individual in the starting population. The number of descendants of each of these ancestors at some time $t$ defines its family size. In the BD process, since all individuals are independent, the evolution of the size of a family behaves the same way as the size of a population (i.e., the number of surviving lineages) that started with a single individual.

 When focusing on the fate of a single ancestor, both the BD and the \SFV~processes have one absorbing state: Whenever a family size reaches zero, the processes stay in that state (the family has become ``extinct''). The BD processes that correspond to \SFV~processes are either critical ($\lambda=\mu$, which is the case we consider here) or supercritical ($\lambda > \mu$). In the critical BD, when starting with one individual at time $0$, the process will eventually reach $0$ with probability one, i.e., any family will become extinct after a sufficient amount of time (although the expected time to that event is infinite). Starting from one individual at time $0$ and conditioning on non-extinction, the probability $p^*_{1m}(t)$ of observing a certain family size $m>0$ at time $t$ for the critical process is given by
 \begin{equation}\label{eq:bdfsizes}
     p^*_{1m}(t)=\frac{(\lambda t)^{m-1}}{(1+\lambda t)^m}
 \end{equation}
as stated in \cite{tavare2018} (Equation 4). It is noteworthy that $p^*_{1m}(t)$ converges in distribution if and only if the process is subcritical ($\lambda<\mu$) \cite{karlin1975,cavender1978}, which the \SFV~is unable to emulate.

Since birth and death rates in the \SFV*~correspond to a critical BD, the family size of any individual from the initial population evolving under \SFV~is expected to drop to $0$ after some (potentially infinite) time. On the other hand, while the death rates are constant and the same for all individuals in the \SFV, the birth rate of one individual may be affected by the number of individuals close by; for example, if a neighbourhood $C\subseteq\HAB$ is momentarily sparsely populated, the probability of a specific individual located in $C$ to be chosen as the ancestor in a birth event is slightly elevated. This spatial influence is of course not present in the BD. 

We simulated 100 runs of the \SFV~forward in time with the choices for $\theta$ and $\xi$ as in the previous section such that $\theta^2\xi=1$, and again $\upsilon=1$. We still considered a rectangle of size $10\times10$ and the population density was set to $\rho=4$. Under these assumptions, at time $0$ the number of individuals on the rectangle is Poisson-distributed with mean $400$ (i.e., $\rho w h$). After $T=1$ and $T=4$ units of simulated time, we recorded the distribution of family sizes and formed the average over all runs. The same was done for a BD with $\lambda=\mu=2\pi$.
\begin{figure}
    \centering
    \includegraphics[width=0.68\textwidth]{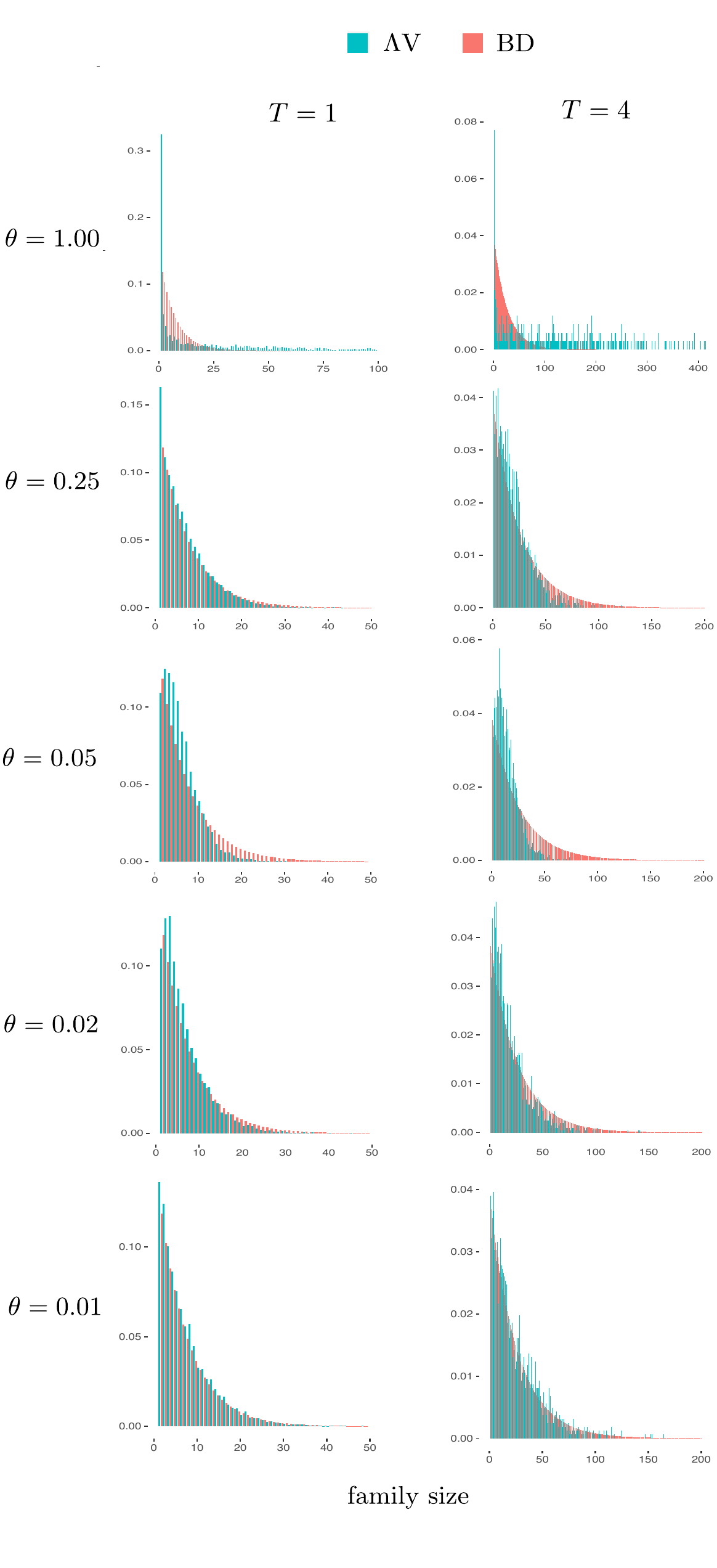}
    \caption{{\bf Distribution of family sizes under BD and \SFV~for $T=1$ (left column) and $T=4$ (right column) for various radii}. The distribution were obtained analytically for BD (see main text) and simulated forward in time for \SFV.}
    \label{fig:famsize}
\end{figure}

The frequencies of family sizes under the BD generally agree well with Eq. \eqref{eq:bdfsizes}. 
Hence we represent the BD by this function in Figure~\ref{fig:famsize}. For the \SFV~process, the absolute values of $\theta$ and $\xi$ visibly affect the shape of the distribution. If $\theta$ is large and events comparably rare (e.g., in the setting $\theta=1$, $\xi=1$), we observe an overabundance of families of size one, and a much flatter distribution otherwise, with extended frequencies of higher family sizes. This observation is most likely explained by the variance in offspring number when a REX event takes place, which is given by $2\pi\theta^2\rho\upsilon$ and thus quadratic with respect to $\theta$. Smaller values of $\theta$ typically provide a good fit between \SFV~and BD. However, after four units of time, we observe a deficit in large family sizes in \SFV~versus BD. That discrepancy probably reflects the impact of spatial constraints in the \SFV. Indeed, families with most members located close to a boundary give birth to a smaller number of individuals compared to those located far away from these boundaries. This difference of behaviour is probably responsible, at least in part, for the observed divergence between the two models although additional investigations are clearly needed in order to have a deeper understanding of the forces at play.

\subsection{Properties of BD and \SFV$^*$ as tree-generating processes}

The statistical properties the BD model as a tree-generating process are well-known, e.g., with respect to branch lengths and tree topology (see e.g., \cite{gernhard2006}). In particular, consider the following setting: Assume that the BD process is initialised at $t_{\text{or}}$ units of time in the past with a single lineage, and there are $n>0$ lineages alive at the present (time $t_0=0$). Consider the joint density 
\begin{equation}
p_{\text{BD}}(\tau,t_1,\dots,t_{n-1} \mid n,\lambda,\mu,t_0,t_{\text{or}})    
\end{equation}
of the topology $\tau$ and the bifurcation times $T_1,\dots,T_{n-1}$ in the past of the family genealogy (where $T_1$ is the most recent bifurcation and $T_i<T_{i+1}$), given the family size $n$, the time frame $[0,t_{\text{or}}]$ and the parameters of the process. Then, it holds that \begin{equation}\label{eq:jointtreedensity}
    p_{\text{BD}}(\tau,t_1,\dots,t_{n-1} \mid n,\lambda,\mu,t_0,t_{\text{or}})\propto \prod_{i=1}^{n-1} p_1(t_i)
\end{equation}
where $p_1(t)$ is the probability that a BD process starting at time $0$ with one lineage has again one single lineage after $t$ units of time (see e.g., \cite{yang1997}). For a critical BD process ($\lambda=\mu$), we have
\begin{equation}
    p_1(t)=\frac{1}{(1+\lambda t)^2}
\end{equation}

We compare $p_{\text{BD}}(\PHY,t_1,\ldots,t_{n-1} \mid \lambda,\mu,t_0,t_{\text{or}},n)$ to $p_{\text{\SFV}}(\PHY, t_1,\ldots,t_{n-1} \mid \xi,\theta,t_0,t_{\text{or}},n)$ through simulations in the case where $n=2$.
We generated trees under the \SFV~process forward in time using the following procedure: the value of $t_{\text{or}}$ is chosen arbitrarily and the corresponding initial location is chosen uniformly at random in $\HAB$. We run the process, updating the genealogy of descendants of the founder after each REX event, until time $0$ is reached. Simulations are discarded whenever the number of lineages $n$ is different from two. We retain a sample of genealogies with valid realisations of $T_1$.  We then compared the empirical distribution of this random variable to that derived analytically for the BD. We repeated these simulations for different values of $\theta$, with $\xi$ chosen such that $\xi=1/\theta^2$, and therefore $\lambda=2\pi$. We opted for $t_{\text{or}}=0.5$, since this suffices to outline the shape of $T_1$ for the range of values of $\theta$ selected here.

\begin{figure}
    \centering
    \includegraphics[width=0.8\textwidth]{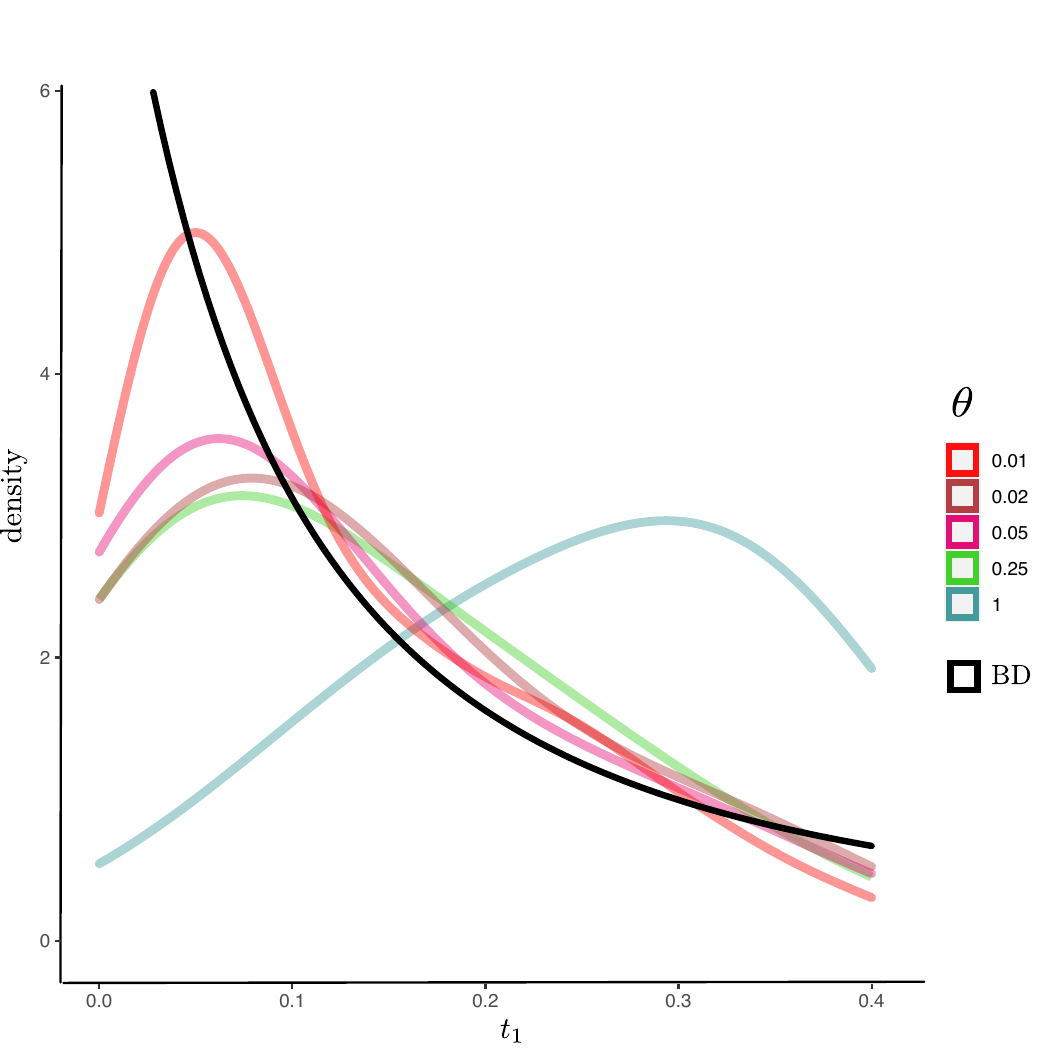}
    \caption{{\bf Distributions of $T_1$ for two-tip trees under the \SFV~and the BD processes.} The distributions for \SFV~were obtained from simulations with 100 repeats for each value of $\theta$ while that for the BD (in black) is analytical (see main text).}
    \label{fig:t1-fsize2}
\end{figure}

Figure \ref{fig:t1-fsize2} shows that for large radii ($\theta=1$, in particular), the distribution of coalescence times of two lineages noticeably diverge in shape and mode from that derived from the BD process. We hypothesise that the number of REX events involved in these particular simulation settings is relatively small so that lineages have to ‘‘wait" relatively long periods of time before being affected by an event, preventing early coalescent events. For smaller values of $\theta$ (and therefore larger values of $\xi$), distributions of $T_1$ derived from the \SFV~are more similar to that given by the BD, as expected.

We now focus on $n=3$ and compare the bifurcation times $T_1$ and $T_2$ obtained under the \SFV~and the BD processes. 
For the joint density of the split times in a critical BD conditioned on $n=3$ and $t_{\text{or}}$, it holds that 
\begin{equation}
    p_{\text{BD}}(\tau,t_1,t_2 \mid n=3,\lambda,\mu,t_0,t_{\text{or}})\propto  \frac{1}{(1+\lambda t_1)^2} \cdot \frac{1}{(1+\lambda t_2)^2}
\end{equation}
Since $t_1<t_2$, we can calculate the marginal density $p_{\text{BD}}(t_1 \mid n,\lambda,t_0,t_b)$ for $T_1$ under the above conditions and using \eqref{eq:jointtreedensity}:
\begin{align}\label{eq:mdensity-t1}
p_{\text{BD}}(t_1 \mid n,\lambda,t_0,t_{\text{or}})&\propto p_1(t_1)\int_{t_1}^{t_{\text{or}}}p_1(t_2)\mathrm{d}t_2\\
&=\frac{1}{(1+\lambda t_1)^2}\cdot\left[\frac{1}{\lambda+\lambda^2  t_1}-\frac{1}{\lambda+\lambda^2 t_{\text{or}}}\right]
\end{align}
Similarly, we obtain the marginal density $p_{\text{BD}}(t_2|n,\lambda,t_{\text{or}})$ of $t_2$:
\begin{align}\label{eq:mdensity-t2}
p_{\text{BD}}(t_2 \mid n,\lambda,t_0,t_{\text{or}})&\propto p_1(t_2)\int_{0}^{t_2}p_1(t_1)\mathrm{d}t_1\\
&=\frac{1}{(1+\lambda t_2)^2}\cdot\left[\frac{1}{\lambda}-\frac{1}{\lambda+\lambda^2 t_2}\right]
\end{align}
As for the \SFV, we repeated the simulations described above, this time discarding all instances where the number of lineages $n$ was not equal to three at time 0. Also, we discarded cases where two of the three final lineages were generated in the same birth event as in such a case $\tau$ is not a binary tree. However, with decreasing $\theta$, this typre of event becomes less and less likely. The bifurcation times $t_1$ and ${t}_2$ can then given by the genealogies obtained from the successful runs. We used the same parameter combinations as in the case of $n=2$, except for $\theta=0.01$ and $\xi=10,000$, as according to our observations, this case becomes numerically infeasible to simulate in a reasonable amount of computing time. Here, the starting point of the simulations was taken as $t_{\text{or}}=2$ units of time in the past.

\begin{figure}
    \centering
    \includegraphics[width=\textwidth]{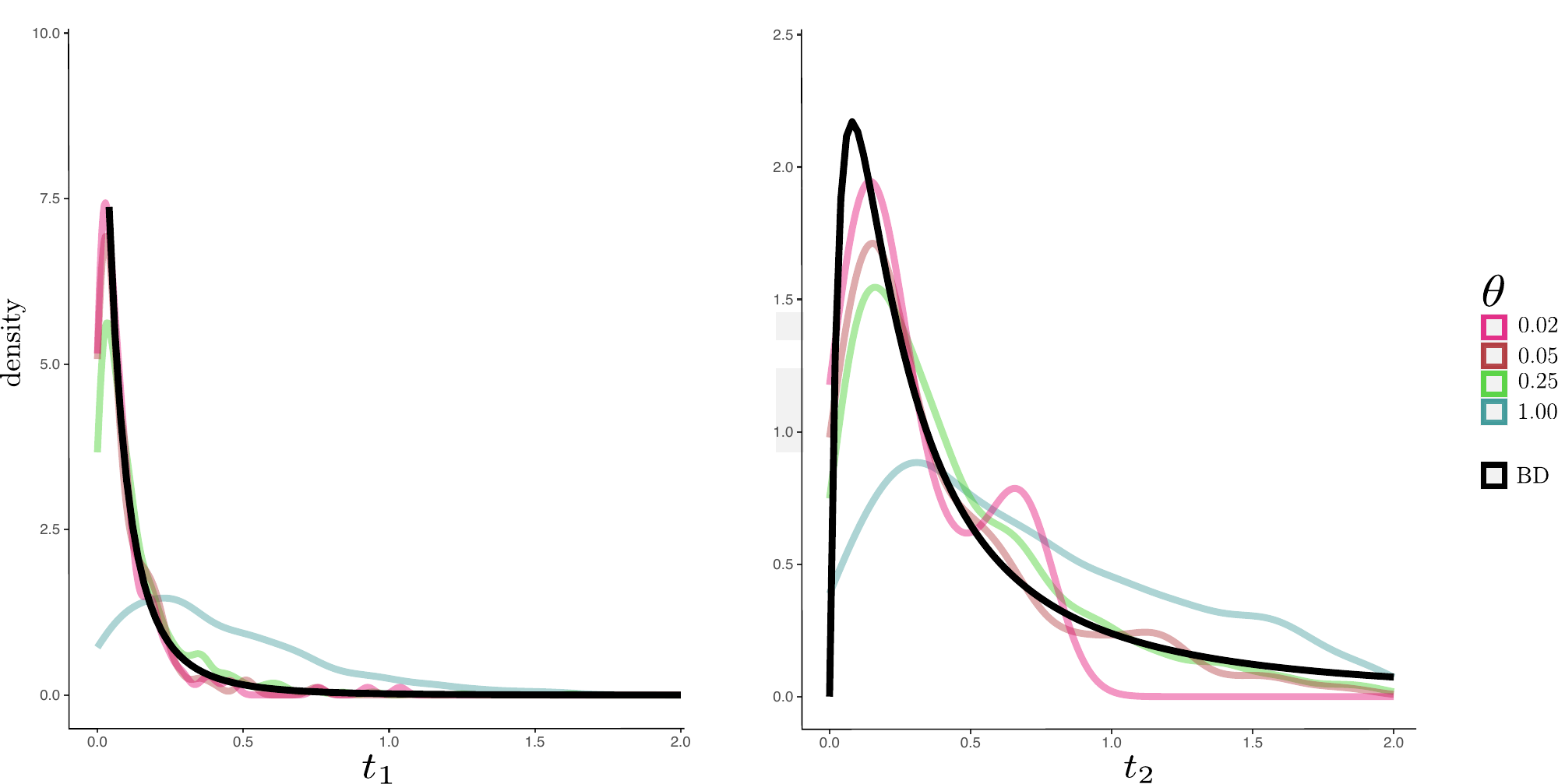}
    \caption{{\bf Distributions of $T_1$ and $T_2$ for three-tip trees under the \SFV~and the BD processes.} The distributions for the \SFV~were obtained from simulations with 100 repeats for each value of $\theta$. The densities corresponding to the BD (in black) agree with Equations \eqref{eq:mdensity-t1} and \eqref{eq:mdensity-t2}.}
    \label{fig:t1t2-fsize3}
\end{figure}

Results in Figure \ref{fig:t1t2-fsize3} indicate a good agreement between distributions of $T_1$ and that of $T_2$ for the two models for values of $\theta$ smaller than 1. Although obtaining a sufficiently large number of valid draws from the target distributions was computationally challenging (hence the rough aspect of some of the curves derived from \SFV~simulations), the modes of the reconstructed densities get closer to that of the BD process when the radius decreases.




\subsection{Comparison of \SFV~and \BDBD~processes}

Results in the previous section indicate that the \SFV~and the BD tree-generating processes are, at least in the simulation settings examined in the present study, equivalent in the limit of a small radius and a large rate of REX events. The present section aims at assessing whether the similarity between the two models still stands when incorporating spatial information. 

When considering a single lineage and ignoring border effects, the movements of the corresponding particle evolving under \SFV~follows a (shifted) Brownian process with diffusion parameter $4\pi\theta^4\xi$ (see Appendix, section \ref{sec:ratesinglelineage}). The behaviour of a pair of lineages is not as straightforward as that of two independent Brownian trajectories. In particular, during the period of time following the birth of the two lineages (i.e., moments after the splitting of their ancestor), the two particles remain in the vicinity of one another. Any given event affecting one of the two particles is thus likely to impact the other as well. The movements of the two particles are therefore not independent and the correlation depends on the time to their common ancestor. Yet, in the limit of a small radius, one may expect the dependency between particles to vanish quickly after their birth and particles may thus be considered as independent when monitored over relatively long periods of time. However, the impact of borders in the habitat can no longer be ignored under the \SFV~while these do not play a role in the \BDBD. The next sections explore these issues using forward and backward in time simulations of two lineages under both processes.


\subsubsection{Comparison of likelihoods}\label{sec:like_comp}

We first focus on the comparison of both models by considering their respective predictions of the spatial coordinates at the tips of a two-lineage tree with fixed ancestral node age and location. The density of interest is noted here as $q_f(L_2,L_3 \mid t_1, l_1, \theta, \xi, \upsilon, h, w, n=2)$, corresponding to the joint density of coordinates $L_2$ and $L_3$ at the tips of lineages 2 and 3, given the time $t_1$ at which these two lineages coalesce, $l_1$ the location of the ancestor just before the edge splitting event and the parameters of the \SFV~model (with $\lambda=\mu=2\pi$ and $\sigma^2=4\pi\theta^2c$ for \BDBD). The subscript $f$ in the density stands for ``forward in time''. 

The habitat is modelled as a $10 \times 10$ square (i.e., $h=w=10$) with an ancestral location set to $l_1=(5,5)$. The radius $\theta$ is fixed to 0.025 throughout these simulations and the rate of events $\xi$ is equal to $1/\theta^2$ so that $c=1$, as per usual. We then obtained the joint distributions of $L_2$ and $L_3$ for values of $t_1$ equal to 100, 1,000 and 5,000. 

\begin{figure}
    \centering
    \includegraphics[width=\textwidth]{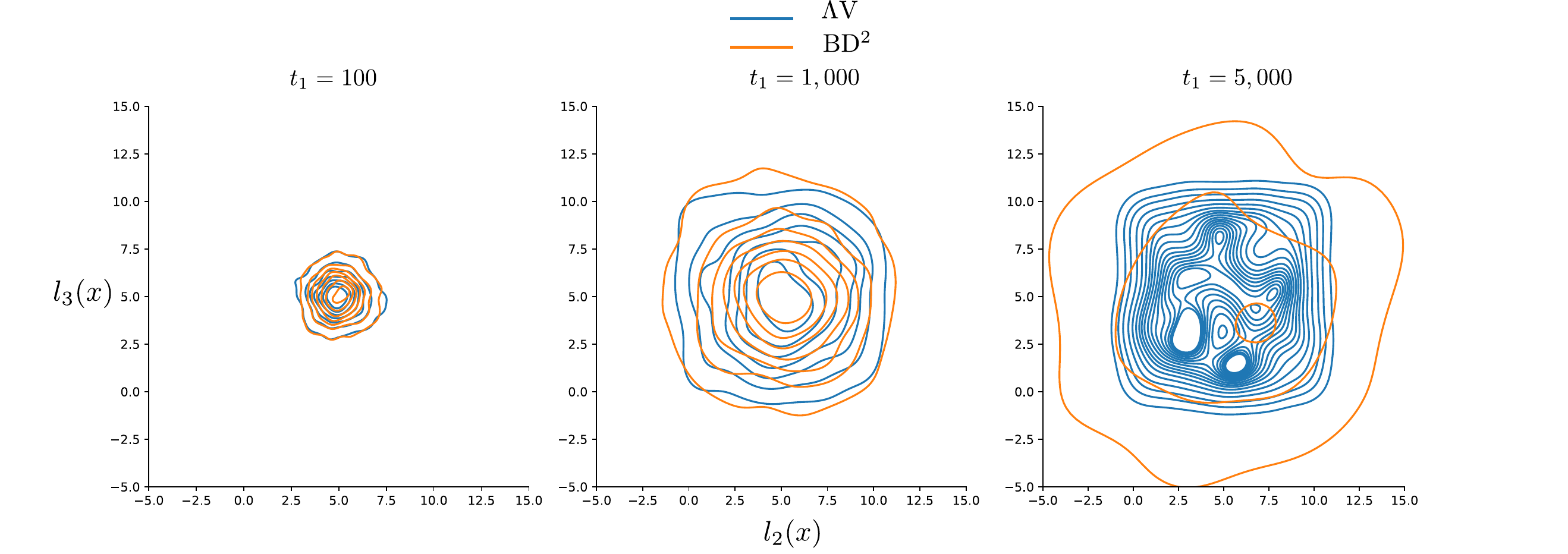}
    \caption{{\bf Distributions of tip locations under the \SFV~and \BDBD~models.} For each model, we generated 1,000 draws from the corresponding distribution with density $q_f(L_2,L_3 \mid t_1, l_1, \theta, \xi, \upsilon, h, w, n=2)$, for values of $t_1=100$ (left), 1,000 (centre) and 5,000 (right), with $\theta=0.025$ and $\xi=1/\theta^2$. The density plots display the joint distributions of $L_2$ and $L_3$ along the $x$-axis (denoted $l_2(x)$ and $l_3(x)$ respectively). Both axes have lower and upper limits -5.0 and +15, to be compared with limits of the habitat (i.e., lower and upper limits of 0 and 10 along both axes)}
    \label{fig:likeallt}
\end{figure}

Figure \ref{fig:likeallt} shows that for relatively small values of the coalescence time, both models predict virtually identical distributions of locations at the tips. In other words, tip locations under the \SFV~are well approximated by a multivariate normal when the radius of events is small compared to the size of the habitat and the rate of REX events is large. For larger values of $t_1$, border effects impact \SFV~substantially and the \BDBD~model puts a large probability mass on tip locations falling outside the habitat (see Figure \ref{fig:likeallt} right). In these conditions, the distribution of $L_2$ and $L_3$ under \SFV~becomes almost uniform and is thus clearly distinct from a bivariate normal (even in the case where realisations of $L_2$ and $L_3$ under \BDBD~are generated using a bivariate normal truncated to $[0,h] \times [0,w]$ so as to better accommodate for the limits of the habitat (results not shown)).

\subsubsection{Comparison of posterior densities}

Results obtained in section \ref{sec:like_comp} indicate that the likelihood of both models are only equivalent in cases where the time to coalescence is not too distant in the past so that the impact of the limits of the habitat can be safely ignored. We now focus on the distribution of the coalescence time and the corresponding ancestral location conditioned on the sampled locations of the two focal lineages. Let $q_b(L_1,T_1 \mid l_2, l_3, \theta, \xi, \upsilon)$ denote that distribution, with the subscript $b$ for the ``backward in time'' process. The forward ($q_f(\cdot)$, see previous section) and backward $(q_b(\cdot)$) distributions bare obvious connections (see below). Yet, just because recent coalesence times most likely generate pairs of tips that are located in a small area (see Figure \ref{fig:likeallt} left) does not necessarily imply that the most probable times of coalescence of lineages sampled in such region are young. 

\begin{figure}
    \centering
    \includegraphics[width=\textwidth]{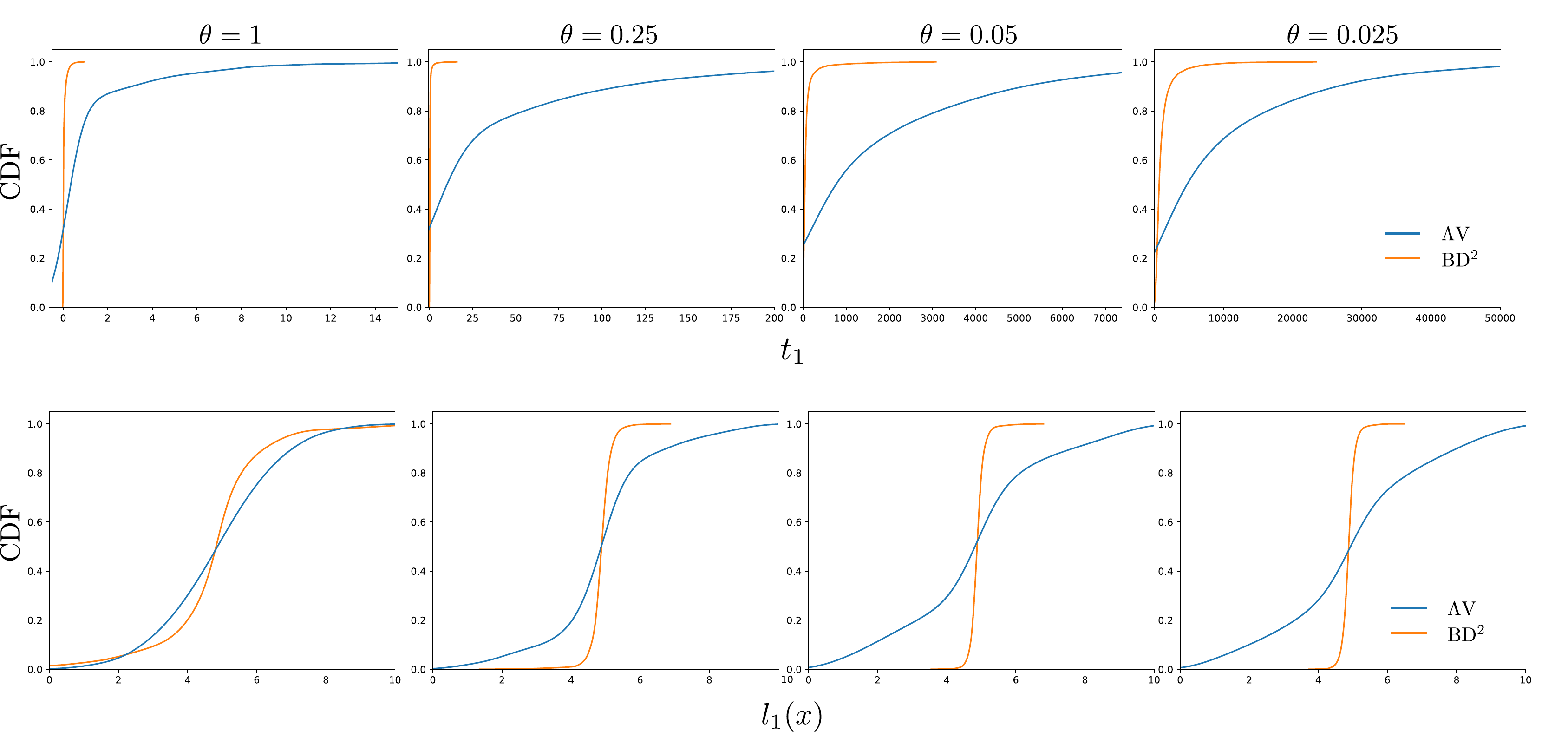}
    \caption{{\bf Posterior distribution functions of coalescent times (top row) and spatial coordinates (bottom row) under the \SFV~and \BDBD~models.} Samples from the distribution with density $q_b(L_1,T_1 \mid l_2, l_3, \theta, \xi, \upsilon)$ were generated under both models for values of $\theta=$1, 0.25, 0.05 and 0.025 (with $\xi=1/\theta^2$ and $\upsilon=1$). Tip coordinates for the two sampled lineages were set to (5.00,5.43) and (4.75, 5.00). }
    \label{fig:postallrad}
\end{figure}

We generated samples from the target distribution through direct simulation under the \SFV~model (see section \ref{sec:fast_back_sim}). As for the \BDBD~model, we obtained correlated samples by applying a Metropolis-Hastings algorithm \cite{metropolis1953,hastings1970}  with standard proposal operators for updating the time to coalescence and the corresponding spatial coordinates. Figure \ref{fig:postallrad} shows the distribution function of $T_1$ and $L_1$ (focusing on the $x$-axis) obtained under the two models for tip coordinates set to $l_2=(5.00,5.43)$ and $l_3=(4.75,5.00)$, and habitat size defined using $w=h=10$. We considered a similar range of values for the radius as the one used previously, i.e., $\theta=$ 1, 0.25, 0.05 and 0.025. The distribution of $T_1$ shows a behaviour similar to that observed when ignoring spatial information with cumulative distributions of the two models becoming more similar as the radius decreases. Results obtained for the spatial component of the models are noticeably different. The range of values for $L_1$ is much narrower under \BDBD~compared to \SFV, with a distribution converging to coordinates generally tightly grouped midway between the two sampled tip locations (i.e., $(5.00+4.75)/2$ along the $x$-axis), while the \SFV~shows a much broader distribution of estimated ancestral locations, even though an inflexion of the distribution function is observed as well around the midpoint between the sampled lineages.

At first glance, the comparison of results obtained by running \BDBD~forward and backward may appear puzzling: forward in time simulations show considerable variance in tip coordinates (see Figure \ref{fig:likeallt}, right) while backward in time simulations, starting from the most likely tip locations and considering time to coalescent of the same order of magnitude, yields very precise coordinates at the coalescent node (see Figure \ref{fig:postallrad}, right). For a fixed coalescent time $t_1$, the posterior distribution of the spatial coordinates at the coalescent node is derived as follows:
\begin{eqnarray*}
    q_b(l_1 \mid l_2, l_3, t_1, \sigma) &\propto& q_f(l_2, l_3 \mid l_1, t_1, \sigma) \\
    &\propto& \phi(l_2;l_1,\sigma^2 t_1) \phi(l_3;l_1,\sigma^2 t_1) \\
    &=& \phi(l_1; \frac{l_2+l_3}{2}, \frac{\sigma^2 t_1}{2})
\end{eqnarray*}
where $\phi(\cdot,\mu,\sigma^2)$ is the normal density with mean $\mu$ and variance $\sigma^2$. Although the following statement lacks a sound mathematical backing, one may argue that the diffusion parameter of the backward in time process is thus half that of the forward in time process.  This observation explains, at least partially, the difference of behaviour of the forward and backward versions of the \BDBD. Explaining the differences between the \BDBD~and the \SFV~models is less straightforward. Conditioned on the time to coalescence of the two focal lineages, 
the distribution of REX events is no longer uniform in space, prohibiting simple mathematical results about the spatial coordinates on the ancestor. The simulation results presented in this study simply suggest that, when focusing on the spatial component of the models, the \SFV~and \BDBD~behave differently in the limit of small radius and frequent REX events even though both models are equivalent when focusing on a single lineage.

\subsection{Efficient simulations under the SLFV model}

Comparison between the \SFV~and other tree- and spatial coordinates-generating processes depends on our ability to efficiently simulate data under these stochastic models. In particular, for the \SFV~model, the present study required simulation under the backward and forward in time versions of this process. As seen above, the forward process generates realisations that may be used for direct comparison with the likelihood of the model. Backward simulations are used instead for comparison with the posterior densities of ancestral node ages and their spatial coordinates. 

Since we focus on the limit of small radius in the present study, the vast majority of REX events do not hit any lineage, making the simulations computationally inefficient. For instance, the backward generation of a two-lineage data set with $\theta=0.02$ takes about 45 minutes for lineages that are 0.5 space unity away from each other on a 10 $\times$ 10 square. Also, naive forward in time simulations require to monitor the whole population of lineages and keep track of their positions at each REX event, which is costly in terms of memory usage. We provide below two algorithms for forward and backward simulation of two lineages evolving under the \SFV~process that alleviate these difficulties.

\subsubsection{Forward simulations}

Our objective here is to obtain independent random draws from the distribution with density $q_f(L_2,L_3 \mid t_1, l_1, \theta, \xi, \upsilon, n=2)$. In words, we want to generate  locations $L_2$ and $L_3$ for two focal lineages (2 and 3, sampled at time 0) given that their most recent common ancestor split at time $t_1$ and had location $l_1$ just before the split. In the following, we first give an algorithm that simulates the trajectory of two lineages forward in time which does not require monitoring the whole population. We then describe a modified, more efficient version of this method that ignores events that leave the two lineages unchanged.

We first generate the position $z$ of the REX event corresponding to the split of the lineage ancestral to 2 and 3 by sampling from a truncated normal distribution with mean $l_1$, variance $\theta^2$ and truncation set so that $z$ falls within the $h \times w$ rectangle defining the habitat. Next, we choose the initial position of each of the two focal lineages, noted $l_2$ and $l_3$, by sampling from a truncated normal with mean $z$ and variance $\theta^2$. (1) The time to the next REX event is then obtained by sampling from an exponential distribution with rate $\xi w h $. (2) The position of that event is selected uniformly at random in the $h \times w$ rectangle. (3) The probability that the sampled lineage $i$ is hit by this event is $u_i(z) = \upsilon \exp\left(-\frac{||l_i-z||^2}{\theta^2}\right)$ (noted as $u_i$ in the following), where $i=2~\text{or}~3$. Also, the probability that both lineages are hit is $u^* = u_1 u_2$. We need to exclude the situation where both lineages are hit by the REX event, since only one parent is selected to give birth to new lineages per REX event. If both were hit by one event, at least one of the two lineages would die without offspring and would therefore not survive to the present time ($t=0$). (4) The probability that one and only one of the two lineages is hit is thus $u_1 + u_2 - 2u^*$. If this event takes place, it affects lineage $i$ with probability $\frac{u_i-u^*}{u_1+u_2-2u^*}$ and the new position of lineage $i$ is sampled from a truncated normal with mean $c$ and variance $\theta$. Steps (1)-(4) of the above procedure are repeated until the time elapsed, i.e. the sum of exponentially distributed times generated in (1), exceeds $t_1$.

The present study focuses on the case where the radius of events is small compared to the size of the habitat. As already mentioned, in this situation, most events do not impact any of the sampled lineages, conveying limited information for our purpose (the rate of these events enters the model as a time scaling factor). We thus elected to adapt our simulation procedure so as to focus solely on the rate of events where one sampled lineage and only one dies. This rate is simply the product of the rate of all events ($\xi h w$) by the probability that one of the two lineages dies, i.e., $\frac{1}{hw} \int_{z \in \mathcal{A}} (u_1+u_2-2u^*) \mathrm{d}z$, where $\mathcal{A}$ is the $h \times w$ rectangle and therefore varies with the lineages' positions (see Appendix for the solution to that integral).

When focusing only on events that impact the sampled lineages, the spatial position of the event centres is no longer uniform. Deriving the joint distribution of the REX centre position along with that of the two lineages right after the event is thus essential in designing an approach that generates random draws from the correct distribution. Although the ordering in which lineages are considered when examining the impact of an event is not relevant, we hereby consider our two focal lineages in a serial fashion, i.e., one lineage is considered as the first while the other is the second. Let $H_1$ be a discrete random variable with state space $\{2,3\}$ corresponding to the event space \{``lineage 2 is the first lineage and dies'',``lineage 3 is the first lineage and dies''\}. Also, let $(H_2 \mid z)$ be the random variable with state space $\{1,2\}$ corresponding to the event space  $\{\text{``the second lineage dies''}, \text{``the second lineage does not die''}\}$. The probability density of interest is thus noted as: 
\begin{align*}
p(H_1 = 2, H_2 = 2, &~z \mid \theta) + p(H_1 = 3, H_2 = 2, z \mid \theta) \\
=& \Pr(H_1 = 2) p(z \mid H_1 = 2, \theta) \Pr(H_2 = 2 \mid z, H_1 = 2) + \\ 
& \Pr(H_1 = 3) p(z \mid H_1 = 3, \theta) \Pr(H_2 = 2 \mid z, H_1 = 3) \\
=& \frac{1}{2} p(z \mid H_1 = 2, \theta) \Pr(H_2 = 2 \mid z) + \\ 
& \frac{1}{2} p(z \mid H_1 = 3, \theta) \Pr(H_2 = 2 \mid z)
\end{align*}
Examination of the last expression suggests that the following procedure could be used in order to get a valid random draw for the event centre according to the model of interest: (1) pick one of the two lineages uniformly at random as the first lineage. Let $i$ denote the event corresponding to the death of that lineage; (2) sample the value of $z$ from the distribution with density $p(z \mid H_1 = i, \theta)$; (3) let $u$ be a random draw from $U[0,1]$, if $u \leq \Pr(H_2 = 2 \mid z, H_1 = i)$ (i.e., the second lineage dies), return to (1), otherwise return $z$.
 
\subsubsection{Backward simulations}\label{sec:fast_back_sim}

The goal of the backward simulations is to generate independent random draws from the distribution with density $q(T_1,L_1 \mid l_2, l_3, \theta, \xi, \upsilon)$, i.e., given $l_2$ and $l_3$, the locations of the two sampled lineages at present, generate $T_1$ and $L_1$, the time and location of their most recent common ancestor. As noted above, if done naively, simulation of the \SFV~when tracking a small number of sampled lineage is computationally costly since the vast majority of REX events do not impact any of the sampled lineage. A more efficient approach would then be to focus exclusively on the REX events that either hit one lineage only, or hit both of them as is the case when coalescence take place, and set the rate of these events in an appropriate manner. Below is a description of one such approach.

The rate of events that hit one or the two lineages is given by the product of the rate of all types of events ($\xi w h$) by the probability that one or the two lineages are hit, i.e., using the notation from the previous section: $\frac{1}{wh} \int_{z \in \mathcal{A}} (u_1 + u_2 - u_1u_2) \mathrm{d}z$ (see section \ref{sec:2linhit} in the Appendix). Hence, here again, this rate is not constant in time as it changes with the position of lineages. The core of the proposed procedure relies on the distribution of the location of a REX event conditioned on that event hitting both lineages or only one of them. Using a similar approach as for the forward case, let $H_1$ be a discrete random variable with state space $\{2,3\}$ corresponding to the event space \{``lineage 2 is the first lineage and is hit'', ``lineage 3 is the first lineage and is hit''\}. Also, let $(H_2 \mid z)$ be the random variable with state space $\{1\}$ corresponding to the event space  \{``the second lineage is hit or not''\}. The joint probability density of one or the two lineages being hit by the event and the location of the REX event is thus expressed as follows:
\begin{align*}
    p(H_1=2,H_2=1,&z \mid \theta) + p(H_1=3,H_2=1,z \mid \theta) \\  
    =&\Pr(H_1=2) \times p(z \mid H_1=2, \theta) \times \Pr(H_2=1|z,H_1=2) +  \\
    & \Pr(H_1=3) \times p(z \mid H_1=3, \theta) \times \Pr(H_2=1|z,H_1=3)\\
    =& \Pr(H_1=2) \times p(z \mid H_1=2, \theta) + \Pr(H_1=3) \times p(z \mid H_1=3, \theta)
\end{align*}
The last expression above suggests that simulating a valid value for the centre position can be done by first picking one of the lineages to be hit by the event with probability $\Pr(H_1=\cdot)=1/2$ and then sampling the event centre from a truncated normal centred on that lineage (with variance $\theta^2$), i.e., with the corresponding density $p(z \mid H_1=\cdot, \theta)$. The simulation continues if the second lineage is not hit by the same event. It stops if the second lineage is hit by the event. In the second case, one then samples $L_1$ from a truncated normal centred on $z$ and the simulation is complete.

\section{Discussion}

The present study illustrates several parallels between the \SFV~and BD models. Starting from the observation that in the \SFV~lineages experience birth and death events over the course of time in a manner similar to the BD, we derived analytical results concerning the rates of these events in the \SFV~when approaching the limit $\theta\rightarrow 0$,  $\xi\rightarrow\infty$ and $\theta^2\xi\rightarrow c$ for constant $c$ (we use \SFV* to denote this particular version of the \SFV). We verified through simulations the theoretical predictions and investigated several related questions regarding the genealogical process in the \SFV. The \SFV~was simulated backward and forward in time in accordance with its standard formulation \cite{barton2010,barton2013}. We introduced two algorithms that permit efficient simulation by skipping REX events that do not impact the sampled lineages. We also implemented forward and backward numerical techniques, through direct simulation or the sampling of correlated samples through MCMC, under the BD tree-generating process and BD with Brownian evolution of spatial coordinates along the tree edges (the so-called $\text{BD}^2$ model). 

We first focused on the tree generating process induced by the \SFV* model. Our simulations indicate that the per lineage birth and death rates do indeed converge to that derived analytically, thereby establishing a first connection with the BD model. We next focused on the distribution of the number of descendants of individual lineages after fixed amounts of time. Here again, we observe a good agreement between the two models, especially for short waiting times. The distributions become distinct for longer time periods, at which point the size of surviving families is large so that the effect of the limited size of the habitat cannot be ignored under the \SFV~model, while it plays no role under the BD~model. Finally, forward simulations suggest that the times to first and second coalescent events in samples of size three in the \SFV~converge in distribution to those observed in the BD. Altogether, our results indicate that the tree-generating processes induced by the \SFV*~and BD processes are equivalent as long as the sample size is small enough so that the limits of the habitat can safely be ignored.

When spatial coordinates of lineages are taken into account, the finite rectangle we simulate on with the \SFV~process induces boundary effects, causing a differentiation between the densities of ancestral lineage locations for the \SFV~and the \BDBD~models. This discrepancy does not vanish with larger rates of REX events and smaller radius. Here, the impact of a decreasing radius does not seem to be offset by the increasing rate of events, pushing coalescent times deeper in the past, thus making the probability for any lineage to hit the habitat boundaries before coalescing non negligible. Backward in time simulations of the dynamics of a pair of lineages show that the spatial distribution of the most recent common ancestor is substantially less variable under \BDBD~compared to \SFV*. This observation entails serious consequences in practice as it implies that the choice of model will impact on the precision with which ancestral coordinates are to be estimated, with \BDBD~potentially giving overly precise estimates when the model that actually generated the data of interest is closer to \SFV.

Finally, we present two algorithms for simulating the temporal and spatial dynamics of a pair of lineages forward and backward in time. These new methods are computationally efficient as they focus solely on REX events that impact the lineages under scrutiny while the naive approach simulates vast numbers of events affecting individuals in the population that are not incorporated in the sample. Importantly, the new backward in time algorithm may serve as a basis for the simulation-based inference of model parameters under the \SFV~(using, for instance, approximate Bayesian computation). While the \SFV~model is amenable to parameter inference \cite{guindon2016}, the task is computationally challenging. Efficient approximation for the time to coalescence of pairs of lineages were derived recently \cite{wirtz2022}. Yet, fast and accurate parameter estimation methods are still lacking and the proposed simulation algorithm presented in this study may contribute to filling this void.

\section*{Acknowledgements}
This work was financially supported by the Agence Nationale
pour la Recherche [\url{https://anr.fr/}] through the grant GENOSPACE, and the Walter-Benjamin Program (WI 5589/1-1) of the DFG [\url{https://dfg.de/}].

\section{Appendix}

\subsection{Dispersal of a single lineage under \SFV}\label{sec:ratesinglelineage}

When considering the  backward in time \SFV~process, the rate at which  a lineage is hit by a  REX is the
product of  the rate  at which  these events  occur  ($\xi w h = \xi|\mathcal{A}|$)  by the
probability that  a lineage is hit.  Let $l^+$  be the (two-dimensional vector) location of the focal lineage
just  before  the REX  event  that  occurred  at time  $t$.   The probability that  this lineage is hit
conditional on the REX event having location $z=(z_x,z_y)$ is
\begin{align}
\int \frac{u(l^+,z)}{|\mathcal{A}|} \mathrm{d}^2l^+ &= \int_0^h \int_0^w \frac{u(l^+,z)}{|\mathcal{A}|}
\mathrm{d}l_x \mathrm{d}l_y \\  
= \frac{\pi \upsilon \theta^2}{2|\mathcal{A}|}
&\left[\text{erf}\left(\frac{\sqrt{2}z_x}{2\theta}\right)-\text{erf}\left(\frac{\sqrt{2}(z_x-w)}{2\theta}\right)\right]
\nonumber \\
\times & \left[\text{erf}\left(\frac{\sqrt{2}z_y}{2\theta}\right)-\text{erf}\left(\frac{\sqrt{2}(z_y-h)}{2\theta}\right)\right].
\end{align}
In cases where the argument of each error function above is large enough (i.e., greater than $\simeq 2$),
its value is close to one. These conditions are met when $\theta \ll \min(z_x,z_y)$
and $z$ is far enough from the edges of the habitat (i.e., $w-z_x \gg 0$ and $z_y \gg 0$, and likewise for $z_y$) . In this situation, the expression above
simplifies, yielding
\begin{eqnarray}
\int \frac{u(l^+,z)}{|\mathcal{A}|} \mathrm{d}^2l^+ \simeq 2\pi \upsilon  \theta^2/|\mathcal{A}|,
\end{eqnarray}
which is also the marginal probability of the lineage being hit (i.e., without conditioning on the position of the REX event).
We will consider that this approximation holds in what follows. The rate at which a given lineage is
hit  is thus  $2\xi \pi  \theta^2 \upsilon$.   

Also, the  probability density  of $l^-$  (the position of the lineage just after the REX event, still going backward in time) given
$l^+$  (with $l^-\neq l^+$)  is $\frac{1}{4\pi^2  \theta^4}\int v(l^-,z_i)
  v(l^+,z_i)   \mathrm{d}^2   z_i$.   This   integral   yields   $\frac{1}{4\pi\theta^2}
\exp\left(-\frac{1}{4\theta^2}||l^- - l^+||^2\right)$,   i.e.,  a  bivariate  normal
density with mean $l^+$ and covariance  matrix $2\theta^2 \mathbf{I}$. The variance of
offspring location  in a one-dimensional space given the parental location  is thus $\Ex(d^2_x)=2\theta^2$;
where $d^2_x$ is the squared Euclidean distance in a one dimensional habitat. $\theta^2$ is thus half the expected square Euclidean distance between parent and offspring in one dimension.
In two dimensions, we have $\Ex\left(\frac{1}{2}(d_x^2 + d_y^2)\right) = \frac{1}{2}\left(\Ex(d_x^2) + \Ex(d_y^2)\right) = 2\theta^2$
i.e., $\theta^2$ is a quarter of the  expected square Euclidean distance between parent and
offspring. In a $n$-dimensional habitat, $\theta^2$ is $1/2n$  times this  expected distance.

Altogether, in a two-dimensional habitat, the variance of spatial
coordinates of a lineage along a given axis thus increases with time proportionally to $\sigma^2 := 4\theta^4  \xi \pi \upsilon$. In  the limit where $\lambda \to \infty$ and $\theta  \to 0$, we hypothesise that the backward-in-time
  motion of a single lineage is a Brownian process with diffusion parameter $\sigma^2$.

\subsection{Probability of coalescence of two lineages}\label{sec:2linhit}

Let $l_2=(l_{2,x},l_{2,y})$ and $l_3=(l_{3,x},l_{3,y})$ be the current positions of the two lineages under scrutiny. The probability that lineage $i$ is hit given the centre position $c=(c_x,c_y)$ is, by definition of the \SFV~model, $u_i(c) = \upsilon \exp\left(-\frac{||l_i-c||^2}{2\theta^2}\right)$. 
The probability that both lineages are hit (i.e., coalesce) is obtained following an approach similar to that used for a single lineage (see above):
\begin{align*}
     \Pr(\text{lineages 2 and 3 are hit}) \\
=& \frac{\upsilon^2}{|\mathcal{A}|} \int_0^w \int_0^h \exp\left( -\frac{(l_{2,x}-z_x)^2 +
    (l_{2,y}-z_y)^2}{2\theta^2} \right) \times \\  
    & \quad \quad \quad \quad \quad \exp\left( -\frac{(l_{3,x}-z_x)^2 +
   (l_{3,y}-z_y)^2}{2\theta^2} \right) \mathrm{d}z_x \mathrm{d}z_y \\
  =& \frac{\pi \theta^2\upsilon^2}{4|\mathcal{A}|} \exp\left(-\frac{(l_{2,x}-l_{3,x})^2 +
     (l_{2,y}-l_{3,y})^2}{4\theta^2}\right) \times \\
  & \Bigg(\text{erf}\left(\frac{l_{2,x}+l_{3,x}}{2\theta}\right) -
     \text{erf}\left(\frac{l_{2,x}+l_{3,x}-2w}{2\theta}\right) \Bigg) \\     
   & \Bigg(\text{erf}\left(\frac{l_{2,y}+l_{3,y}}{2\theta}\right) - \text{erf}\left(\frac{l_{2,y}+l_{3,y}-2h}{2\theta}\right) \Bigg)
\end{align*}
and the probability of coalescence gets close to $\frac{\pi \theta^2\upsilon^2 }{|\mathcal{A}|} \exp\left(-\frac{||l_2-l_3||^2}{4\theta^2}\right) $ for small values of $\theta$.

\bibliographystyle{elsarticle-harv}
\bibliography{ref}

\begin{thebibliography}{32}
\expandafter\ifx\csname natexlab\endcsname\relax\def\natexlab#1{#1}\fi
\providecommand{\url}[1]{\texttt{#1}}
\providecommand{\href}[2]{#2}
\providecommand{\path}[1]{#1}
\providecommand{\DOIprefix}{doi:}
\providecommand{\ArXivprefix}{arXiv:}
\providecommand{\URLprefix}{URL: }
\providecommand{\Pubmedprefix}{pmid:}
\providecommand{\doi}[1]{\href{http://dx.doi.org/#1}{\path{#1}}}
\providecommand{\Pubmed}[1]{\href{pmid:#1}{\path{#1}}}
\providecommand{\bibinfo}[2]{#2}
\ifx\xfnm\relax \def\xfnm[#1]{\unskip,\space#1}\fi
\bibitem[{Barton et~al.(2010)Barton, Etheridge and V{\'e}ber}]{barton2010}
\bibinfo{author}{Barton, N.}, \bibinfo{author}{Etheridge, A.},
  \bibinfo{author}{V{\'e}ber, A.}, \bibinfo{year}{2010}.
\newblock \bibinfo{title}{A new model for evolution in a spatial continuum}.
\newblock \bibinfo{journal}{Electronic Journal of Probability}
  \bibinfo{volume}{15}.
\bibitem[{Barton et~al.(2013)Barton, Etheridge and V{\'e}ber}]{barton2013}
\bibinfo{author}{Barton, N.H.}, \bibinfo{author}{Etheridge, A.M.},
  \bibinfo{author}{V{\'e}ber, A.}, \bibinfo{year}{2013}.
\newblock \bibinfo{title}{Modelling evolution in a spatial continuum}.
\newblock \bibinfo{journal}{Journal of Statistical Mechanics: Theory and
  Experiment} \bibinfo{volume}{38}, \bibinfo{pages}{P01002}.
\bibitem[{Biswas et~al.(2021)Biswas, Etheridge and Klimek}]{biswas2021}
\bibinfo{author}{Biswas, N.}, \bibinfo{author}{Etheridge, A.},
  \bibinfo{author}{Klimek, A.}, \bibinfo{year}{2021}.
\newblock \bibinfo{title}{The spatial lambda-fleming-viot process with
  fluctuating selection}.
\newblock \bibinfo{journal}{Electron. J. Probab.} \bibinfo{volume}{26},
  \bibinfo{pages}{1--51}.
\bibitem[{Bradburd and Ralph(2019)}]{bradburd2019}
\bibinfo{author}{Bradburd, G.S.}, \bibinfo{author}{Ralph, P.L.},
  \bibinfo{year}{2019}.
\newblock \bibinfo{title}{Spatial population genetics: it's about time}.
\newblock \bibinfo{journal}{Annual Review of Ecology, Evolution, and
  Systematics} \bibinfo{volume}{50}, \bibinfo{pages}{427--449}.
\bibitem[{Bradburd et~al.(2016)Bradburd, Ralph and Coop}]{bradburd2016}
\bibinfo{author}{Bradburd, G.S.}, \bibinfo{author}{Ralph, P.L.},
  \bibinfo{author}{Coop, G.M.}, \bibinfo{year}{2016}.
\newblock \bibinfo{title}{A spatial framework for understanding population
  structure and admixture}.
\newblock \bibinfo{journal}{PLoS genetics} \bibinfo{volume}{12},
  \bibinfo{pages}{e1005703}.
\bibitem[{Cavender(1978)}]{cavender1978}
\bibinfo{author}{Cavender, J.A.}, \bibinfo{year}{1978}.
\newblock \bibinfo{title}{Quasi-stationary distributions of birth-and-death
  processes}.
\newblock \bibinfo{journal}{Advances in Applied Probability}
  \bibinfo{volume}{10}, \bibinfo{pages}{570--586}.
\bibitem[{Drummond and Rambaut(2007)}]{drummond2007}
\bibinfo{author}{Drummond, A.J.}, \bibinfo{author}{Rambaut, A.},
  \bibinfo{year}{2007}.
\newblock \bibinfo{title}{{BEAST}: {B}ayesian evolutionary analysis by sampling
  trees}.
\newblock \bibinfo{journal}{BMC Evolutionary Biology} \bibinfo{volume}{7},
  \bibinfo{pages}{214}.
\bibitem[{Felsenstein(1975)}]{felsenstein1975}
\bibinfo{author}{Felsenstein, J.}, \bibinfo{year}{1975}.
\newblock \bibinfo{title}{A pain in the torus: some difficulties with models of
  isolation by distance}.
\newblock \bibinfo{journal}{American Naturalist} \bibinfo{volume}{109},
  \bibinfo{pages}{359--368}.
\bibitem[{Gernhard(2006)}]{gernhard2006}
\bibinfo{author}{Gernhard, T.}, \bibinfo{year}{2006}.
\newblock \bibinfo{title}{Stochastic models for speciation events in
  phylogenetic trees}.
\newblock \bibinfo{journal}{arXiv preprint math/0610919} .
\bibitem[{Guindon et~al.(2016)Guindon, Guo and Welch}]{guindon2016}
\bibinfo{author}{Guindon, S.}, \bibinfo{author}{Guo, H.},
  \bibinfo{author}{Welch, D.}, \bibinfo{year}{2016}.
\newblock \bibinfo{title}{Demographic inference under the coalescent in a
  spatial continuum}.
\newblock \bibinfo{journal}{Theoretical Population Biology}
  \bibinfo{volume}{111}, \bibinfo{pages}{43--50}.
\bibitem[{Hastings(1970)}]{hastings1970}
\bibinfo{author}{Hastings, W.K.}, \bibinfo{year}{1970}.
\newblock \bibinfo{title}{Monte carlo sampling methods using markov chains and
  their applications} .
\bibitem[{Joseph et~al.(2016)Joseph, Hickerson and
  Alvarado-Serrano}]{joseph2016}
\bibinfo{author}{Joseph, T.}, \bibinfo{author}{Hickerson, M.},
  \bibinfo{author}{Alvarado-Serrano, D.}, \bibinfo{year}{2016}.
\newblock \bibinfo{title}{Demographic inference under a spatially continuous
  coalescent model}.
\newblock \bibinfo{journal}{Heredity} \bibinfo{volume}{117},
  \bibinfo{pages}{94--99}.
\bibitem[{Karlin and Taylor(1975)}]{karlin1975}
\bibinfo{author}{Karlin, S.}, \bibinfo{author}{Taylor, H.M.},
  \bibinfo{year}{1975}.
\newblock \bibinfo{title}{Chapter 9 - stationary processes}, in:
  \bibinfo{editor}{Karlin, S.}, \bibinfo{editor}{Taylor, H.M.} (Eds.),
  \bibinfo{booktitle}{A First Course in Stochastic Processes (Second Edition)}.
  \bibinfo{edition}{second edition} ed.. \bibinfo{publisher}{Academic Press},
  \bibinfo{address}{Boston}, pp. \bibinfo{pages}{443--535}.
\bibitem[{Kimura(1953)}]{kimura1953}
\bibinfo{author}{Kimura, M.}, \bibinfo{year}{1953}.
\newblock \bibinfo{title}{`{S}tepping stone' model of population}.
\newblock \bibinfo{journal}{Annual Report of the National Institute of Genetics
  Japan} \bibinfo{volume}{3}, \bibinfo{pages}{62--63}.
\bibitem[{Kingman(1982)}]{kingman1982}
\bibinfo{author}{Kingman, J.F.C.}, \bibinfo{year}{1982}.
\newblock \bibinfo{title}{On the genealogy of large populations}.
\newblock \bibinfo{journal}{Journal of Applied Probability}
  \bibinfo{volume}{19(A)}, \bibinfo{pages}{27--43}.
\newblock \DOIprefix\doi{10.2307/3213548}.
\bibitem[{Lemey et~al.(2010)Lemey, Rambaut, Welch and Suchard}]{lemey2010}
\bibinfo{author}{Lemey, P.}, \bibinfo{author}{Rambaut, A.},
  \bibinfo{author}{Welch, J.J.}, \bibinfo{author}{Suchard, M.A.},
  \bibinfo{year}{2010}.
\newblock \bibinfo{title}{Phylogeography takes a relaxed random walk in
  continuous space and time}.
\newblock \bibinfo{journal}{Molecular Biology and Evolution}
  \bibinfo{volume}{27}, \bibinfo{pages}{1877--1885}.
\bibitem[{Lemmon and Lemmon(2008)}]{lemmon2008}
\bibinfo{author}{Lemmon, A.R.}, \bibinfo{author}{Lemmon, E.M.},
  \bibinfo{year}{2008}.
\newblock \bibinfo{title}{A likelihood framework for estimating phylogeographic
  history on a continuous landscape}.
\newblock \bibinfo{journal}{Systematic Biology} \bibinfo{volume}{57},
  \bibinfo{pages}{544--561}.
\bibitem[{Louvet(2023)}]{louvet2023}
\bibinfo{author}{Louvet, A.}, \bibinfo{year}{2023}.
\newblock \bibinfo{title}{Stochastic measure-valued models for populations
  expanding in a continuum}.
\newblock \bibinfo{journal}{ESAIM: Probability and Statistics}
  \bibinfo{volume}{27}, \bibinfo{pages}{221--277}.
\bibitem[{Mal{\'e}cot(1948)}]{malecot1948}
\bibinfo{author}{Mal{\'e}cot, G.}, \bibinfo{year}{1948}.
\newblock \bibinfo{title}{Mathematics of heredity}.
\newblock \bibinfo{publisher}{Paris: Masson et Cie.}
\bibitem[{Metropolis et~al.(1953)Metropolis, Rosenbluth, Rosenbluth, Teller and
  Teller}]{metropolis1953}
\bibinfo{author}{Metropolis, N.}, \bibinfo{author}{Rosenbluth, A.W.},
  \bibinfo{author}{Rosenbluth, M.N.}, \bibinfo{author}{Teller, A.H.},
  \bibinfo{author}{Teller, E.}, \bibinfo{year}{1953}.
\newblock \bibinfo{title}{Equation of state calculations by fast computing
  machines}.
\newblock \bibinfo{journal}{The Journal of Chemical Physics}
  \bibinfo{volume}{21}, \bibinfo{pages}{1087--1092}.
\bibitem[{Novembre et~al.(2008)Novembre, Johnson, Bryc, Kutalik, Boyko, Auton,
  Indap, King, Bergmann, Nelson et~al.}]{novembre2008}
\bibinfo{author}{Novembre, J.}, \bibinfo{author}{Johnson, T.},
  \bibinfo{author}{Bryc, K.}, \bibinfo{author}{Kutalik, Z.},
  \bibinfo{author}{Boyko, A.R.}, \bibinfo{author}{Auton, A.},
  \bibinfo{author}{Indap, A.}, \bibinfo{author}{King, K.S.},
  \bibinfo{author}{Bergmann, S.}, \bibinfo{author}{Nelson, M.R.}, et~al.,
  \bibinfo{year}{2008}.
\newblock \bibinfo{title}{Genes mirror geography within europe}.
\newblock \bibinfo{journal}{Nature} \bibinfo{volume}{456},
  \bibinfo{pages}{98--101}.
\bibitem[{Rousset(2003)}]{rousset2001}
\bibinfo{author}{Rousset, F.}, \bibinfo{year}{2003}.
\newblock \bibinfo{title}{Inferences from spatial population genetics}, in:
  \bibinfo{editor}{Balding, D.}, \bibinfo{editor}{Bishop, M.},
  \bibinfo{editor}{Cannings, C.} (Eds.), \bibinfo{booktitle}{Handbook of
  statistical genetics}. \bibinfo{publisher}{Wiley}.
\bibitem[{Suchard et~al.(2018)Suchard, Lemey, Baele, Ayres, Drummond and
  Rambaut}]{suchard2018}
\bibinfo{author}{Suchard, M.A.}, \bibinfo{author}{Lemey, P.},
  \bibinfo{author}{Baele, G.}, \bibinfo{author}{Ayres, D.L.},
  \bibinfo{author}{Drummond, A.J.}, \bibinfo{author}{Rambaut, A.},
  \bibinfo{year}{2018}.
\newblock \bibinfo{title}{Bayesian phylogenetic and phylodynamic data
  integration using {BEAST} 1.10}.
\newblock \bibinfo{journal}{Virus Evolution} \bibinfo{volume}{4},
  \bibinfo{pages}{vey016}.
\bibitem[{Tavar{\'e}(2018)}]{tavare2018}
\bibinfo{author}{Tavar{\'e}, S.}, \bibinfo{year}{2018}.
\newblock \bibinfo{title}{The linear birth--death process: an inferential
  retrospective}.
\newblock \bibinfo{journal}{Advances in Applied Probability}
  \bibinfo{volume}{50}, \bibinfo{pages}{253--269}.
\bibitem[{Véber and Wakolbinger(2015)}]{veber2015}
\bibinfo{author}{Véber, A.}, \bibinfo{author}{Wakolbinger, A.},
  \bibinfo{year}{2015}.
\newblock \bibinfo{title}{The spatial {L}ambda-{F}leming-{V}iot process: An
  event-based construction and a lookdown representation}, in:
  \bibinfo{booktitle}{Annales de l'IHP Probabilit{\'e}s et statistiques}, pp.
  \bibinfo{pages}{570--598}.
\bibitem[{Wang et~al.(2012)Wang, Z{\"o}llner and Rosenberg}]{wang2012}
\bibinfo{author}{Wang, C.}, \bibinfo{author}{Z{\"o}llner, S.},
  \bibinfo{author}{Rosenberg, N.A.}, \bibinfo{year}{2012}.
\newblock \bibinfo{title}{A quantitative comparison of the similarity between
  genes and geography in worldwide human populations} .
\bibitem[{Wilkins(2004)}]{wilkins2004}
\bibinfo{author}{Wilkins, J.F.}, \bibinfo{year}{2004}.
\newblock \bibinfo{title}{A separation-of-timescales approach to the coalescent
  in a continuous population}.
\newblock \bibinfo{journal}{Genetics} \bibinfo{volume}{168},
  \bibinfo{pages}{2227--2244}.
\bibitem[{Wilkins and Wakeley(2002)}]{wilkins2002}
\bibinfo{author}{Wilkins, J.F.}, \bibinfo{author}{Wakeley, J.},
  \bibinfo{year}{2002}.
\newblock \bibinfo{title}{The coalescent in a continuous, finite, linear
  population}.
\newblock \bibinfo{journal}{Genetics} \bibinfo{volume}{161},
  \bibinfo{pages}{873--888}.
\bibitem[{Wirtz and Guindon(2022)}]{wirtz2022}
\bibinfo{author}{Wirtz, J.}, \bibinfo{author}{Guindon, S.},
  \bibinfo{year}{2022}.
\newblock \bibinfo{title}{Rate of coalescence of lineage pairs in the spatial
  $\lambda$-{F}leming--{V}iot process}.
\newblock \bibinfo{journal}{Theoretical Population Biology}
  \bibinfo{volume}{146}, \bibinfo{pages}{15--28}.
\bibitem[{Wright(1931)}]{wright1931}
\bibinfo{author}{Wright, S.}, \bibinfo{year}{1931}.
\newblock \bibinfo{title}{Evolution in {M}endelian populations}.
\newblock \bibinfo{journal}{Genetics} \bibinfo{volume}{16},
  \bibinfo{pages}{97}.
\bibitem[{Wright(1943)}]{wright1943}
\bibinfo{author}{Wright, S.}, \bibinfo{year}{1943}.
\newblock \bibinfo{title}{Isolation by distance}.
\newblock \bibinfo{journal}{Genetics} \bibinfo{volume}{28},
  \bibinfo{pages}{114}.
\bibitem[{Yang and Rannala(1997)}]{yang1997}
\bibinfo{author}{Yang, Z.}, \bibinfo{author}{Rannala, B.},
  \bibinfo{year}{1997}.
\newblock \bibinfo{title}{Bayesian phylogenetic inference using dna sequences:
  a markov chain monte carlo method.}
\newblock \bibinfo{journal}{Molecular biology and evolution}
  \bibinfo{volume}{14}, \bibinfo{pages}{717--724}.

\end{thebibliography}
\end{document}